# Optical Tests of a 2.5-m diameter Liquid Mirror: Behavior under External Perturbations and Scattered Light Measurements

Luc Girard, and Ermanno F. Borra,
Centre d'Optique, Photonique et Laser,
Département de Physique, Université Laval, Québec, Qc, G1K 7P4
email : borra@phy.ulaval.ca, girard@phy.ulaval.ca

astro-ph/9706030   3 Jun 1997

## Abstract

Interferometric tests of a f/1.2 2.5-m diameter liquid mirror show RMS surface deviations ~ $\lambda/20$ and Strehl ratios of order 0.6, showing that it is diffraction limited. The mirror is certainly better than implied by the data because of aberrations introduced by the auxiliary testing optics. We have made detailed studies of the scattered light of the mirror. We have studied the behavior of the mirror under external perturbations. With this article we have reached an important milestone since we now have a good understanding of liquid mirrors.

## 1. Introduction

Following the suggestion [1] that modern technology renders liquid mirrors useful to astronomy, a feasibility study was undertaken to determine whether, in practice, it is possible to generate an optical quality surface on a spinning liquid. This led to optical tests that showed that a 1.5-m diameter liquid mirror (LM) had such good optical quality that it was diffraction-limited [2]. The article [2] gives a wealth of information on the basic LM technology. This was followed by tests of a 2.5-m diameter liquid mirror [3] that also showed diffraction limited performance. An important milestone has been reached by Content et al. [4] who published the first scientific paper that reported astronomical research with a liquid mirror telescope (LMT). Deep-sky mages have been taken with a LMT that tracked with a driftscanning CCD [5].

Liquid mirrors are interesting in other areas of science besides astronomy. For example, the University of Western Ontario has built a Lidar facility that houses a 2.65-m diameter liquid mirror as a receiver [6]. A Lidar facility has also been built and operated by the University of California at Los Angeles [7]. Liquid mirrors are also used as reference surfaces to test conventional optics [8].

In this article we discuss the final results of extensive tests of the 2.5-m LM that complete the earlier ones reported in reference [3]. We discuss scattered light measurements and the behavior of the mirror subjected to misalignments and external perturbations. A more detailed description of the tests can be found in Girard's Ph.D. thesis[9]. A review paper [10] gives a convenient summary of the status of LMTs and related issues.

## 2. Wavefronts

The basic mirror setup and testing facilities are essentially the same as in [2] and [3]. We have two new custom made null lenses designed by C. Morbey that reimage the mirror to f/8.4. Most of the interferometry is done with a scatter plate interferometer [12] but some is done with a Shack-cube interferometer [13]. The interferograms are captured with 1/30 second exposure times by a 512X480 CCD detector connected to an 8-bit framegrabber interfaced with a PC/AT clone computer. They are analyzed with software that uses a Fourier technique [14] capable of giving a substantially greater resolution and signal-to-noise ratio than the usual fringe-following algorithms.

Figure 1 shows a typical interferogram and Figure 2 shows the wavefront obtained from it. The spatial resolution on the mirror is typically 6X6 cm but we can get resolutions as high as 1.5X4 cm. As discussed in [2] and [3], we remove from the raw wavefronts mean values of focus, coma and spherical aberrations. The mean values are small and are obtained from averages of several wavefronts. Experiments and computer simulations give us compelling reasons to believe that they are produced by the auxiliary optical components in the testing setup as discussed in previous articles [2,3]. Independent measurements of a 1.4-m liquid mirror have been made [8], finding negligible coma and spherical aberration within the uncertainties of the measurements.

In this section we present the results of interferometric tests of well-tuned mirrors in different conditions (e.g. different mercury thicknesses), as well as mirrors that have been perturbed with the express purpose of studying their behaviors under external perturbations and misalignments. We also study spiral-shaped defects that are induced by the interaction of the rotating surface of the mirror with the air.

2.1 Dependence on layer thickness

We have obtained wavefronts for several thicknesses of mercury. We cannot overemphasize the importance of working with thin mercury layers, not only because they reduce the weight borne by the mechanical structures but, most importantly, because they dampen better any disturbance [2]. On the other hand, a thin layer requires a more precise top surface of the container and introduces greater print-through [2], resulting in more scattered light, as discussed in section 3.

Table 1 summarizes the results of the analysis of 8 interferograms taken consecutively over a 10-minute time interval for a layer of mercury 1.5-mm thick. The interferograms were taken at random, but at about 1 minute intervals. The 1/30 second capture times are sufficiently short that we can detect rapid liquid movements, but they also render the interferometry sensitive to seeing effects in the testing tower. We also obtained similar interferograms for thicknesses of 0.85 mm, 2.0 mm and 3.0 mm. Table 2 summarizes the statistics of the wavefronts taken at different thicknesses. It gives mean values for various statistics of individual wavefronts, as well as statistics of average wavefronts. The mean values and average wavefronts are obtained from 8 wavefronts for each thickness. Prima facie, it would appear that the 2-mm layer gives marginally better individual wavefronts. However, in practice, the difference is probably not significant. Because of the time it takes to change the layer thickness and the time it takes the mirror to stabilize, the measurements were taken on different days; hence the mirror and the null lenses had to be retuned. As a consequence, we feel that the differences between thicknesses seen in Table 2 reflect differences in tuning and alignments rather than thickness-related differences. On the other hand, Table 2 shows a systematic increase of the Strehl ratio with decreasing thickness for the average wavefronts. The average wavefront is less sensitive to alignment-induced time variations, hence the increase probably reflects an actual increase in optical quality with decreasing layer thickness as one would expect, given the greater damping with thinner layers.

2.2 Aberrations of well-tuned mirrors

Visual inspection of the wavefronts indicates that most of the deviations from a perfect wavefront come from the edges. To quantify the contributions of the edges, we analyze the mirror with the edges masked at various diameters. Table 3 shows the Strehl ratios, averaged over all thicknesses, for different diaphragms. We can see a significant increase of the Strehl ratio with decreasing diaphragm size.

We have investigated the causes that degrade the Strehl ratio. This was done by extensive analysis of the behavior of the main aberrations with time and other physical parameters. The detailed analysis can be found in [9]. We do find that poorly tuned mirrors have the same kind of surface aberrations that are described in [2]. However, for a well-tuned mirror, we find that most of the aberrations are not on the mirror but are caused by small tilt and focal length time variations of the mirror. These are negligible by themselves, however they introduce misalignments of the mirror with respect of the null lenses which result in significant aberrations. The mirror suffers from small variations of its rotational velocity as well as of the inclination of its axis of symmetry, the latter caused by imperfections of the bearing. These effects induce misalignements of the null lenses that vary with the period of rotation of the mirror. Indeed we see variations of coma and tilt that are correlated with the azimuthal angle of the mirror and have its period of rotation. Computer simulations with optical design software that assume known alignment errors predict variations having the observed amplitude. For example, we see focus and spherical aberration variations that are anticorrelated and have a ratio of 1.2, in agreement with numerical simulations. In addition, there is some time varying as well as fixed astigmatism. Interferometric measurements of the auxiliary flat mirrors used in the testing setup show astigmatism of the same order of magnitude as we measure on the wavefronts obtained from the interferograms of the 2.5-m mirror. Computer simulations indicate that an auxiliary quarter wave plate introduces astigmatism, as confirmed by experiments. On the other hand, we cannot totally exclude that some astigmatism having amplitude of about $\lambda/10$ is present on the mirror.

Table 4 gives the Strehl ratios for the average wavefronts of mirrors having different thicknesses and with different diaphragms. These time averages show substantial improvements over the instantaneous values. The Strehl ratios increase with decreasing thickness, as should be expected. They also decrease with increasing diameter of the diaphragm.

2.3 Defects caused by leveling errors

We have investigated the effects that leveling errors have on surface quality. The wavefronts with leveling errors have the same coma-like shapes shown in Figure 10 of [2]. Most of the wavefront error is present at the edges of the mirror. We however found another effect which is more subtle: leveling errors introduce a print-through that scatters light into the wings of the PSF. Defects present on the top surface of the container print-through to the surface of mercury. They can have any spatial frequency, from mm-wide bubbles and scratches to valleys and hills tens of centimeters across. These defects have amplitudes of the order of 0.1 mm on the polyurethane spincast surface of our container. Print-through is caused by perturbations (rotational velocity, leveling errors, etc.) that make the liquid flow on the surface of the container. Figure 3 shows an out of focus PSF of a mirror having a 1 arcsecond leveling error. We can see a cardioid-like feature at the center of the mirror that can be used to adjust the axis of rotation (see the appendix) and also radial defects. These defects are on the surface of the container and are due to the way the container was built, by assembling triangular sections. Because the defects are local, print-through affects little the RMS of the mirror measured with the resolution of the interferometer but introduces scattered light. We find that print-through increases with decreasing mercury thickness.

Table 5 shows the statistics of mean wavefronts for 2 leveling errors and 2 mercury thicknesses. The statistics are shown for full apertures and 90 % apertures. We can see that a 0.5 arcsecond error has a negligible effect that becomes significantly greater for a 1 arcsecond error. The Strehl ratio is worse for the thicker layer, illustrating once more the importance of working with thin mercury layers. We can also see a considerable improvement of the Strehl ratio for the mirror with a diaphragm, as expected from the visual appearances of the wavefronts. The level of a liquid mirror should be adjusted to 0.25 arcsecond for optimal performance. Putting a diaphragm on the mirror improves the Strehl ratio, at the expense of collecting area.

2.4 Spiral-shaped defects

The surface of the mirror shows spiral-shaped defects, hereafter referred to as "spirals". The easiest way to see the spirals is to observe an out of focus PSF (Fig.4). Note that an out of focus PSF can reveal defects having very small amplitudes (~$\lambda$/100). The spirals rotate at a speed that is slightly inferior to the rotational velocity of the turntable. The spirals could be in the air, on the surface of the liquid or in both.

Figure 4 was taken with a layer of mercury 3 mm thick to increase the visibility of the surface spirals. The thicker the layer, the more prominent the spirals and the closer they approach the center of the mirror. For a thickness of 0.5 mm one can only see the spirals up to a distance of about 12 cm from the rim. Note however that stating that the mirror has a thickness of x mm is somewhat misleading. First, the thickness quoted is an average thickness computed from the area of the mirror and the mass of mercury used. Because the mirror sags under the weight of mercury, it is certain that the thickness of mercury is not uniform. Furthermore, the outer 5 cm of the mirror have a greater depth of mercury (6 to 8 mm) needed to stabilize thin mercury layers [2]. Figure 5 shows the outer regions of mirrors having two different thicknesses of mercury (1.8 and 0.8 mm). The scales are matched, but the region from 1.0 m to 1.25 m is missing for the image at 1.8 mm. It shows that the spirals extend much further to the center for the thicker layer. A substantial fraction of the contribution to the spirals obviously comes from the mercury surface because the amplitudes decrease with the thickness of mercury, as expected from liquid damping theory [2]. These observations clearly emphasize the importance of working with as thin a layer of mercury as practical.

We have studied quantitatively the spirals with the scatterplate and the Shack interferometers. Figure 6 summarizes the main information. The figure displays the RMS amplitudes of the spirals measured on the 2.5-m mirror for three thicknesses of mercury as identified on the figures. The data do not extend further inward than 0.8 meters because the amplitudes decreased below detection. Taken at face value Figure 6 indicates that, between 0.85 and 0.95 m, the spirals were stronger for 0.8-mm and 0.5-mm layers than for 1.8-mm layers. This conclusion is mistaken because the measurements were very difficult and, the amplitudes of the spirals being small, the signals tend to be comparable to the noise of the interferometer. Also, the interferometry has a resolution barely sufficient to spatially resolve the spirals. As a consequence, the information contained in Figure 6 must be considered with some caution. One can trust the general trends displayed in the figure, more than the absolute values or detailed features of the curves.

In the same figure, we have plotted the amplitudes of the spirals measured by Content [11] on a 1.5-m liquid mirror. Content made his measurements for thicknesses of mercury that differ from ours but there is enough thickness bracketing to allow us to make a meaningful comparison. Comparing the amplitudes at the rims of the 2 mirrors, we can see that, for a given thickness, the amplitudes of the spirals are greater at the rim of the 2.5-m than they are at the rim of the 1.5-m. This is to be expected since the rim of the 2.5-m travels 1.67 times faster than the rim of the 1.5-m. We can also see that extrapolating the amplitude of the spirals beyond 0.7-m radius and 1.4-mm thickness on the 1.5-m predicts amplitudes that are too large when compared to the data taken with the 2.5-m. This is consistent with the assumption that the spirals are an edge effect. This assumption is also consistent with visual evaluations of the pupils of the 2 mirrors. Comparing images of the pupil of the 1.5-m to those of the 2.5-m, one has the impression that one is looking at similar images; this implies that one pupil can be obtained from the other by multiplying it by the ratio of the diameters. It must be noted that, for technical reasons [2], our mirrors have, at their outer edges, a deeper thickness of mercury ( 6 to 8 mm). Thicker layer have substantially less damping than thin ones [2], one should therefore expect larger amplitude spirals at the very edges even for mirrors that have thin layers everywhere else. One should expect that waves generated at the thick outer edges should travel some distance through the thinner inner layers before being dampened out of detectability.

The Coriolis forces on the rotating turntable tend to impose a spiral structure to any radially moving stream. As a matter of fact, the spirals are caused by spiral-shaped turbulence in the air. Our testing facility has remarkably homogeneous air and little seeing. One may therefore wonder whether liquid mirrors working in a less stable environment may have seeing spirals. To test this hypothesis we have injected above the mirror a weak stream of argon gas. When first injected the argon gas, having a different index of refraction from the air, can be seen, with optical tests, as a turbulent stream superposed on a smooth wavefront. Being heavier than air, it then descends to the surface of the mirror. When the argon stream reaches the boundary layer near the surface of the mirror, it develops a spiral structure, as expected. This simple experiment indicates that a liquid mirror working in an astronomical environment should be expected to have a spiral shaped seeing component. Because the spirals contain high spatial frequency components, they introduce scattered light. To minimize seeing spirals, one should shelter the mirror from air currents. This can readily be done by surrounding it with an enclosure.

2.5 Conclusion of the interferometry

With Strehl ratios of order 0.6, the overall quality of the 2.5-m mirror is excellent, comparable to the quality of the 1.5-m liquid mirror previously tested [2]. Misalignement errors between the mirror and the null lenses, particularly those due to the wobbling of the bearing and varying focal length, induce larger aberrations for the fast f/1.2 2.5-m LM than the slower f/2 1.5-m LM. We have strong reasons to believe that liquid mirrors can be produced having Strehl ratios of order unity. This will require fine tuning of the mirrors, consisting mostly in using thin layers (< 1 mm), controlling the rotational velocity of the turntable and the level of the mirror. Because they have a thicker layer of mercury, the edges of the mirror contribute a disproportionate amount of aberrations, they could be masked for optimum performance. We have not tried very hard to improve our Strehl ratios, since they are more than adequate for Earth-bound observations and, furthermore, the degradation of the Strehl ratios is in large part due to the fact that we measure the mirror through null lenses.

## 3. Scattered Light

We have carried out a detailed study of the scattered light generated by the mirror and its causes. The observations consist of direct observations of the artificial star created by a laser and a spatial filter. We use the same basic instrumental setup of the interferometry, including the null lenses. The measurements were carried out for three distinct mercury thicknesses: 0.54 mm, 0.8 mm and 1.8 mm. Our framegrabber has only 8 bits hence has an insufficient dynamic range to measure an adequate range of the intensity of the PSF, which extends over several orders of magnitude. It is therefore necessary to reconstruct the PSF from series of observations of PSFs having different levels of exposures, some of them with a heavily overexposed core. Figure 7 shows a typical sequence of PSFs. Note that the image in A was taken with a 10X microscope objective. Note also the structure surrounding the heavily saturated PSF at right (Fig. 7.D). This structure is discussed in the remainder of this section. A ghost image is apparent at the right of the PSF in D. It is subtracted during data reduction. Getting the azimuthally averaged composite PSF involves some technically challenging operations described in [2,3]. Because the light beam encounters a total of 23 surfaces, refracting and reflecting, in its journey from the pinhole to the CCD, there is bound to be a significant contribution from the auxiliary optics in the form of scattered light or parasitic reflections.

3.1  Dependence on layer thickness

Figure 8 shows the composite PSFs obtained for the three mercury thicknesses. We can see that scattered light decreases with the thickness of mercury. This decrease quantifies what one sees by looking at the pupil of the mirror revealed by the out-of-focus PSF: that spirals and concentric rings decrease with thickness. The decrease appears particularly important going from 0.8 mm to 0.5 mm. The prominent bump at 12.8 arcseconds, for the 1.8-mm layer is due to the concentric rings seen in Figure 7, while the spirals contribute mostly to the region from 2 to 10 arcseconds. There is some inconsistency for radii $> 15$ arcseconds where the figure seems to indicate that the thicker layer actually has less scattered light than the thinner ones. However, one should not attach too much significance to that region. The far wings are less reliable since difficulties in joining the PSFs and estimatimg the background contribution particularly affect those regions

3.2 Measurements with diaphragmed mirrors

A straight comparison of the three curves in figure 8 underestimates the importance of working with thin mercury layers. To stabilize thin mercury layers, a groove is present around the rim of the container [2]: As a consequence, there always is a thicker layer of mercury at the rim. Although the 5 mm deep groove is present for all thicknesses, it is relatively more important for the thinner layers. As a consequence, damping is not improved that much at the rim by decreasing layer thickness, particularly for the thinner layers. We can see this by looking at an out-of-focus image of the PSF. Figures 9 and 10 quantify this observation. We show three PSFs (Fig. 9) as well as three encircled energy curves (Fig. 10) obtained with a 0.5-mm layer: for the complete mirror, the mirror with a diaphragm that excludes the outer 12% of the radius, as well as the whole mirror with a 0.41 arcsecond leveling error. One can see that the mirror with a diaphragm has substantially less scattered light than the whole mirror. We can also see that a small leveling error does increase the amount of scattered light. This last issue is further discussed in section 3.3.  A similar figure for 0.8 mm [9] shows a similar, but smaller, difference between the mirrors with and without a diaphragm, a result which was expected. The difference is substantially smaller for the 1.8-mm layer [9].

3.3 Scattered light introduced by leveling errors

Figures 9, 10, 11 indicate that leveling errors introduce scattered light. However, tilt-induced misalignments of the null lenses complicate scattered light measurements with leveling errors by affecting the normalization of the PSF. We correct for this effect but there is a significant uncertainty associated with this correction, especially with the 0.41 arcseconds data. To further investigate the effect of a leveling error on scattered light, we have introduced a 1.25 arcsecond leveling error on a mirror having 0.8 mm of mercury. The larger leveling error allows us to discriminate more easily between real leveling-induced scattered light and an erroneous correction for coma. Figure 11 shows four PSFs taken with a 0.8-mm thick layer. Three PSFs were taken consecutively with the same mirror adjustment. Because we were concerned with scattered light measurements that illustrate the effects of the wind and level misalignments, and that, consequently, we are looking for relative changes in scattered light, we did not optimize the painstaking adjustment of the null lenses necessary to obtain a high value of the Strehl ratio. As a consequence the core of the PSF suffers from some null-lens-induced coma and, mostly, spherical aberrations. These aberrations are not on the mirror. The PSF labeled "reference" shows, for comparison, a PSF obtained with good alignments of the null lenses. This is the PSF displayed in Figure 8. The difference between the 2 PSFs at 0.8 mm thickness illustrates the fact that misalignments of the null lenses influence scattered light measurements by affecting the normalization of the PSF. Figure 11 shows that a tilted mirror has excess scattered light. This scattered light comes from a coma-like surface deformation that is caused by a leveling error [2] and from an increased print-through. This coma-like deformation is on the mirror, unlike the coma and spherical aberrations that come from misaligned null lenses. Eye inspection of out of focus PSFs shows that a leveling error introduces more print-through. This is presumably due to the fact that a leveling error causes the mercury layer to slosh over the surface of the container.

To minimize scattered light, the level of the mirror should be adjusted to better than 0.25 arcseconds. Because they have a thicker layer of mercury, the edges of the mirror contribute a disproportionate quantity of scattered light, they could be masked to minimize it.

3.4 Scattered light caused by the wind

We have simulated the effects of a turbulent wind on the mirror by blowing, with a small fan, a horizontal air stream having a speed of a few km/h. An inspection of the out of focus PSF shows an increase in the amplitude of the spirals. We do not know whether these spirals are mostly seeing spirals or surface spirals since the motor of the fan is bound to heat the air stream. The effect of the spirals can readily be seen in Figure 11: the scattered light is increased. Whether the spirals are surface waves or seeing induced, the experiment indicates that one must shelter the mirror from outside air perturbations. This is easily accomplished by sheltering it with a secondary enclosure inside the observatory. We shelter our mirror with a simple polyethylene enclosure. Working with thin mercury layers reduces the amplitude of surface spirals but does not affect the seeing spirals at all.

3.5 Scattered light caused by vibrations (concentric rings)

One of the questions most often asked about liquid mirrors concerns the effects of vibrations. In practice, vibrations were never a serious problem. We work in a building that vibrates and shakes like any building does. The most obvious effect of vibrations can be seen in the shape of concentric waves on the surface of the mirror. These waves can be seen in published knife-edge tests [15] of a 1-m liquid mirror. The concentric waves form a phase grating on the surface of the mirror that gives a ring of light around the PSF. Figure 7.D shows that the ring does not have a clean appearance but has a fair amount of structure. Tapping on the base of the mirror gives a very convenient way to see the effect of vibrations. We then see a wide spectrum of concentric waves resulting in as many as 7 concentric rings around the PSF. One can obtain a measure of the RMS amplitude of the waves from [16]

$$RMS = \sqrt{\frac{E}{E_t}} \lambda / 4\pi \qquad (1)$$

where E is the energy in the first order ring and $E_t$ is the total encircled energy.

We have extracted the contribution from the vibration induced bumps by subtracting the continuum of the PSF. The bumps so extracted are shown in Figure 12 for three mercury thicknesses. Note the decrease in amplitude with decreasing thickness, illustrating once more the importance of working with thin layers. Also note that the location of the bumps changes with mercury thickness, indicating that the resonant frequency of the mirror has changed. This is not surprising since mercury contributes significantly to the mass budget of the mirror. However, the location of the bumps cannot be modeled with a naive harmonic oscillator formula where the mass dependence is given by √mass. Presumably this is due to the complexity of the system that contains an air bearing and a composite material container. It must also be noted that Figure 12 is for a complete mirror. Visual inspection of an out-of focus PSF shows that for the 0.8-mm layer and, especially, the 0.5-mm layer, the concentric waves are much more prominent at the very center of the mirror (masked for the data in Fig. 12) as well as the rim, a behavior expected from the fact that the layer of mercury is thicker at the edges.

The energy contained in the rings is small. The layers having 1.8 mm, 0.8 mm and 0. 5 mm thicknesses contain, respectively, 1.1%, 0.46 % and 0.37% of the total energy. Note that these values refer to the full mirrors. The RMS amplitudes of the surface waves are respectively 0.0082 $\lambda$, 0.0054 $\lambda$ and 0.0049 $\lambda$, however this is an average value for the entire surfaces. The amplitudes of the rings decrease considerably for the thinner layers, and essentially vanish when we put a diaphragm on the mirror eliminating the outer 12 % of the radius. This is expected since the mercury layer is deeper at the rim, resulting in less damping.

In conclusion, vibration-induced concentric rings are not important and are concentrated at the edges of the mirror, where the layer thickness is greater

3.6. Oxide skin

Experiments show that, after a few hours, the surface of mercury develops a transparent skin that decreases substantially the mercury vapors. It has a thickness of about $\lambda/4$. We assume that the skin is mercury oxide. The skin can be broken accidentally, e.g. if there is a sudden change of rotational velocity. Inspection of an out-of-focus PSF shows that, after some accident, there are pieces of overlapping broken skin that introduce small phase discontinuities that will introduce scattered light. A typical incidence of broken skins occurs whenever the mirror is stopped and then restarted without cleaning. For best results, the mirror should be cleaned prior to restarting it. The oxide skin is not a serious problem, especially if the mirror is cleaned every time it is stopped and restarted.

3.7    Dust, dirt and insects

Dust scatters light as it does for a conventional glass mirror. We can also see that mercury puckers around large surface contaminants due to the effect of surface tensions. Fortunately, a liquid mirror is very easy to clean. Insects, on the other hand, can be a bothersome cause of scattered light. As they walk on the surface of mercury, they induce surface ripples. Very large insects can impact the mercury surface with enough energy to poke holes in the liquid. The mirror should be covered with a plastic cover when not operating to keep insects out. During operation, one should use standard devices to eliminate insects (special lights, etc...).

**4.  Conclusion**

Interferometric tests of a liquid mirror having a diameter of 2.5 meters show excellent optical qualities. We find Strehl ratios of about 0.6 for well-tuned mirrors. Most defects (e.g. aberrations, scattered light sources) are at the very rim of the mirror. The Strehl ratios increase significantly if one masks the outer 10% of the mirror. Likewise, scattered light decreases if one masks this outer region. The poorer quality of the outer region is due to the fact that it has a thicker layer of mercury. Most of the deterioration of the wavefronts actually comes from small rotational velocity and level variations of the mirror that are transformed into higher order aberrations by the null lenses: They are not on the surface of the mirror. Our experiments have shown that essentially perfect liquid mirrors can be produced.

With this article we have reached an important milestone since, along with the work in [2], we now have a good understanding of liquid mirrors. Producing a good quality liquid mirror involves more than just spinning a vat filled with mercury. The rotational velocity must be controlled to better than one part in a million and the axis of rotation must be parallel to the gravitational field of the Earth to within 1/4 arcsecond. Throughout this article, we continually emphasize, at the risk of being repetitious, the importance of working with thin layers of mercury. One cannot overemphasize the importance of working with thin layers of mercury because of the effectiveness with which thin layers dampen disturbances. We can see a considerable improvement by going from a 3-mm layer to a 2-mm layer. The thinner the layer, the better the damping. We have produced liquid mirrors with layers as thin as 0.5-mm and thinner layers can be produced but the container must then follow a parabola to a fraction of the thickness of the layer, a feat that becomes increasingly difficult as the layer get thinner. We feel that a 1-mm layer is a reasonable compromise. Note that simply pouring mercury in the container does not allow one to work with a layer thinner than 5 mm. Special techniques are needed, as explained in [2].

This article concentrates on issues related to the optical qualities of mercury mirrors. However there are other important issues about liquid mirrors. The review article by Borra [10] discusses a number of them. In conclusion, although there is room for improvement of this very young technology, we do have a working design that is sufficiently robust to be useful for practical use. We also have an adequate understanding of the behavior of a liquid mirror under perturbations. In other words, we know how to make liquid mirrors that work but one can do better.

# APPENDIX
An optical technique to accurately align the axis of rotation of a liquid mirror.

Measuring leveling errors with a high quality level can certainly be done but, at least with a spirit level, it involves delicate measurements, easily disturbed by the environment.  Furthermore, we find that a level made while the mirror is stopped changes by a few arcseconds after spinup and after the final mercury surface settles. This is presumably due to flexures induced by a nonuniform mercury distribution on the surface of the container. We also find that the level of the mirror changes after pump down [2], probably due to flexure of the container and the mount induced by a nonuniform distribution of mercury. Obviously the level must be checked after the layer is closed. We have found a simple optical test that allows accurate leveling.  The technique uses the grooves present on the surface of the container. The grooves, located at the center and at the edges of the mirror, are needed to allow a thin layer of mercury [2]. The center of the mirror has also a small metal hollow tube to pump down the mercury layer [2].  Because the mercury is deeper in the grooves, a leveling error introduces a local deformation that looks like coma. We found that this coma-like local defect is very sensitive to leveling errors and can be used to align the mirror. The defect is readily visible on out of focus images of the PSF. Figure 13 shows the defect at the center of the mirror that has contours looking like a cardioid curve. The figure shows the appearance of the cardioid for three leveling errors. The shape of the cardioid can be used to level the mirror with an accuracy of 0.1 arcseconds. The cardioid method uses the center of the mirror but the very edges of the mirror are also sensitive to leveling errors and can also be used to adjust the level. However, the edges are not as sensitive to leveling errors as the center of the mirror. Note that there may be an angular shift between the direction of the tilt and the maximum of the coma-like deformation. The shift depends on the thickness of mercury.

ACKNOWLEDGMENTS
We wish to thank Dr. C. Morbey for designing our null lenses. We wish to thank Dr. P. Hickson for showing us how to work with composite materials. Our understanding of the engineering challenges posed by liquid mirrors have greatly benefited from studies carried out by numerous students in the engineering department of Laval University over several years: Dr. C. Gosselin, Dr. L. Cloutier and Mr. P.M. Arsenault helped supervising them. This work has greatly profited from past contributions that Dr. Robert Content has made to the LMT project with software and hardware developments, as well as with numerous useful conversations. This research has been supported by grants from the Natural Sciences and Engineering Research Council of Canada and the Formation des Chercheurs et Aide à la Recherche program of the province of Québec to EFB.

## FIGURE CAPTIONS

FIGURE 1: Typical scatterplate interferogram of the 2.5-m liquid mirror having a mercury layer of 0.85 mm. The shadow of the hollow central cylinder, used for mercury manipulation, can be seen below the characteristic hot spot of the scatterplate interferogram.

FIGURE 2: Wavefront of the mirror with a mercury layer of 0.85 mm. This instantaneous three-dimensional rendering of the surface corresponds to the interferogram shown in figure 1.

FIGURE 3: Defocused PSF of the 2.5-m liquid mirror with a 1-arcsecond leveling error. The cardioid-like feature at the center can be used to adjust the axis of rotation of the mirror. One can also see radial defects caused by movement of the liquid over the container.

FIGURE 4: Defocused PSF of the mirror with a mercury layer of 3.0 mm showing the spiral-shaped defects .

FIGURE 5: Defocused PSF showing the outer region of the mirror with mercury thicknesses of 0.8 and 1.8 mm. The scales are matched so that the edge of the mirror on both images are overlapping. The region from 1.0 m to 1.25 m is missing from the image at 1.8 mm. One can see that the spirals extend much further towards the center of the mirror for the thicker layer.

FIGURE 6: RMS amplitude of the spiral-shaped defects of two mirrors having respectively 1.5-m and 2.5-m diameters. The graph shows that the amplitude is greater for the 2.5-m having a rim speed 1.67 times faster than the 1.5-m. It also indicates that the spirals are an edge effect.

FIGURE 7: Sequence of PSFs used for scattered light measurements. The individual frames are 30 arcseconds wide. One can see in A an unsaturated PSF observed with a 10X microscope objective. (2.5" FOV). In B, C and D, we show the PSF directly imaged on the CCD with various levels of saturation. In D, one see weak structures far from the center of the PSF (concentric ring, diffraction cross, ghost images, etc.).

FIGURE 8: Azimuthally averaged PSFs obtained for three mercury thicknesses on the 2.5-m liquid mirror. They show that scattered light decreases with mercury thickness. The bump at 12.8" for the curve at 1.8 mm is caused by concentric waves on the mirror.

FIGURE 9: Azimuthally averaged PSF obtained with a 0.5 mm mercury layer for the complete mirror, the mirror with a diaphragm that excludes the outer 12% of the radius, as well as the complete mirror with a 0.41" leveling error.

FIGURE 10: Encircled energy curves obtained with a 0.5-mm mercury layer for the entire mirror, the mirror with a diaphragm that excludes the outer 12% of the radius, and the entire mirror with a 0.41" leveling error.

FIGURE 11: Azimuthally averaged PSF obtained with a 0.8-mm mercury layer for the mirror under perturbation. Three curves are obtained with the same null lenses alignments: the complete mirror without external perturbation, the mirror with 1.25" leveling error, and the mirror perturbed with a few km/h wind. The reference for a 0.8 mm mercury thickness, with good null lenses alignments, is also shown.

FIGURE 12: Azimuthally averaged rings extracted from PSFs at various thicknesses of figure 8. These rings are caused by concentric waves on the surface of the mirror acting as a phase grating.

FIGURE 13: Defocused PSFs of the central region of the mirror showing the cardioid-like defect for A) no leveling error, B) 1 arcsec leveling error, and C) 3 arcsecs leveling error.

TABLE 1
WAVEFRONT STATISTICS OF A LIQUID MIRROR
HAVING A 1.5-mm THICK MERCURY LAYER

| Azimuthal angle (deg) | P-V $(\lambda)$[a] | RMS $(\lambda)$[a] | Strehl |
|---|---|---|---|
| 0 | 0.593 | 0.064 | 0.522 |
| 45 | 0.552 | 0.051 | 0.665 |
| 90 | 0.478 | 0.051 | 0.671 |
| 135 | 0.565 | 0.057 | 0.596 |
| 180 | 0.494 | 0.053 | 0.647 |
| 225 | 0.615 | 0.074 | 0.427 |
| 270 | 0.568 | 0.070 | 0.463 |
| 315 | 0.746 | 0.064 | 0.528 |
| Mean values[b] | 0.576 | 0.061 | 0.565 |
| Average wavefront | 0.256 | 0.025 | 0.905 |

[a] Wavelength units of surface deviations: 6328Å
[b] Average statistics of the individual wavefronts

TABLE 2
SUMMARY OF THE WAVEFRONT STATISTICS FOR VARIOUS MERCURY THICKNESSES

| Thickness (mm) | Individual wavefront | | | Average wavefront | | |
|---|---|---|---|---|---|---|
| | <P-V> $(\lambda)$[a] | <RMS> $(\lambda)$[a] | <Strehl> | P-V $(\lambda)$[a] | RMS $(\lambda)$[a] | Strehl |
| 0.85 | 0.553 | 0.074 | 0.448 | 0.239 | 0.025 | 0.907 |
| 1.5  | 0.576 | 0.061 | 0.565 | 0.256 | 0.025 | 0.905 |
| 2.0  | 0.545 | 0.053 | 0.643 | 0.330 | 0.030 | 0.866 |
| 3.0  | 0.626 | 0.072 | 0.451 | 0.378 | 0.042 | 0.756 |

[a] Wavelength units of surface deviations: 6328Å

TABLE 3
AVERAGE STREHL RATIOS WITH
DIFFERENT DIAPHRAGMS
(averaged over all thicknesses)

| Diaphragm[a] | <Strehl> | Std. dev. |
|---|---|---|
| 1.00 | 0.527 | 0.095 |
| 0.95 | 0.650 | 0.092 |
| 0.90 | 0.717 | 0.084 |
| 0.85 | 0.763 | 0.073 |

[a] Effective aperture $r/r_{max}$

TABLE 4
WAVEFRONT STATISTICS FOR THE AVERAGE WAVEFRONTS FOR DIFFERENT THICKNESSES AND DIAPHRAGMS

| Thickness (mm) | 100% | | 95% | | 90% | | 85% | |
|---|---|---|---|---|---|---|---|---|
| | RMS ($\lambda$)[a] | Strehl | RMS ($\lambda$)[a] | Strehl | RMS ($\lambda$)[a] | Strehl | RMS ($\lambda$)[a] | Strehl |
| 0.85 | 0.025 | 0.907 | 0.021 | 0.937 | 0.019 | 0.947 | 0.018 | 0.954 |
| 1.5  | 0.025 | 0.905 | 0.020 | 0.941 | 0.019 | 0.949 | 0.017 | 0.955 |
| 2.0  | 0.030 | 0.866 | 0.026 | 0.902 | 0.022 | 0.928 | 0.020 | 0.941 |
| 3.0  | 0.042 | 0.756 | 0.038 | 0.801 | 0.036 | 0.819 | 0.033 | 0.844 |

[a] Wavelength units of surface deviations: 6328Å

# REFERENCES

1. E.F. Borra,"The Liquid-Mirror Telescope As A Viable Astronomical Tool," Journal of the Royal Astronomical Soc. of Canada 76, 245-256 ( 1982).
2. E.F. Borra, R. Content, L. Girard, S. Szapiel, L.M. Tremblay, & E. Boily, E.F., "Liquid Mirrors: Optical Shop Tests and Contributions to the Technology," Astrophysical Journal 393, 829-847 (1992).
3. E.F. Borra, R. Content, L. Girard, " Optical Shop Tests of a f/1.2 2.5-meter Diameter Liquid Mirror," Astrophysical Journal 418, 943-946 (1993).
4. R. Content, E.F. Borra, M.J. Drinkwater, S. Poirier, E. Poisson, M. Beauchemin, E. Boily, A. Gauthier, & L.M. Tremblay, "A Search for Optical Flashes and Flares with a Liquid Mirror Telescope," Astronomical Journal 97, 917-922 (1989).
5. P. Hickson, E.F. Borra, R. Cabanac, R. Content, B.K. Gibson, & G. A. H. Walker,

TABLE 5
WAVEFRONT STATISTICS FOR THE AVERAGE WAVEFRONTS
FOR 2 TILT ERRORS AND 2 MERCURY THICKNESSES

| Thickness (mm) | Tilt error (arcsec) | 100% | | | 90% | | |
|---|---|---|---|---|---|---|---|
| | | P-V $(\lambda)$[a] | RMS $(\lambda)$[a] | Strehl | P-V $(\lambda)$[a] | RMS $(\lambda)$[a] | Strehl |
| 0.8 | 0.5 | 0.264 | 0.030 | 0.866 | 0.178 | 0.020 | 0.943 |
| 0.8 | 1.0 | 0.720 | 0.077 | 0.397 | 0.321 | 0.050 | 0.675 |
| 1.8 | 1.0 | 1.054 | 0.096 | 0.238 | 0.294 | 0.037 | 0.808 |

[a] Wavelength units of surface deviations: 6328Å

"UBC/LAVAL 2.7-Meter Liquid Mirror Telescope," Astrophysical Journal Letters 436, 201-204 (1994).
6. R. J. Sica, S. Sargoytchev, E.F. Borra, L. Girard, S. Argall, C.T. Sarrow, & S. Flatt, "Lidar Measurements Taken with a Large Aperture Liquid Mirror:1.The Rayleigh-Scatter System," App. Opt., 34, No 30, 6925-6936 ( 1995).
7. R.Wuerker, "Bistatic LMT Lidar Alignment," Opt Eng, to be published March, (1997).
8. N.M. Ninane, & C.A. Jamar, "Parabolic Liquid Mirrors in Optical Shop Testing," Appl. Opt., 35, No 31, 6131-6139 (1996).
9. L. Girard, "Etude d'un miroir liquide de 2.5 mètres de diamètre," Unpublished Ph.D. thesis, Université Laval, Québec, QC, Canada (1997).
10.E.F. Borra, "Liquid Mirrors," Can J. of Phys. 73, 109-125 (1995).
11. R. Content, Tests optiques sur un miroir liquide de 1.5 m et développement de la technologie des miroirs liquides, unpublished Ph.D. Thesis, Université Laval, Québec, QC, Canada (1992).
12. S. Mallick, in Optical Shop Testing, Chapter 3, ed.D. Malacara , (New York: John Wiley & Sons) (1978).
13. R.V. Shack, & G.H. Hopkins, "The Shack Interferometer," Optical Engineering, 18, No 2, 226-228 (1979).
14. C. Roddier, & F. Roddier, "Interferogram Analysis Using Fourier Transform Techniques," App. Opt, 26, 1668-1673 (1987).

15. E. F.Borra, M. Beauchemin, R. Arsenault, & R. Lalande, "Optical-Shop Testing of Liquid Mirrors ," Publications of the Astronomical Society of the Pacific 97, 454-464 (1985).
16. W. B. Wetherell, in Applied Optics and Optical Engineering , Chapter 5, Vol 8, eds. Shannon, R.R., and Wyant, (New York: Academic Press) (1980).

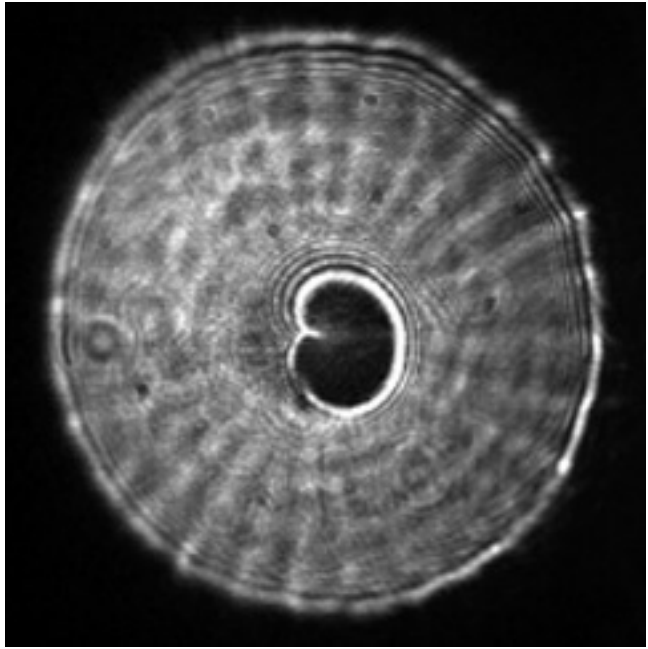

Figure 3

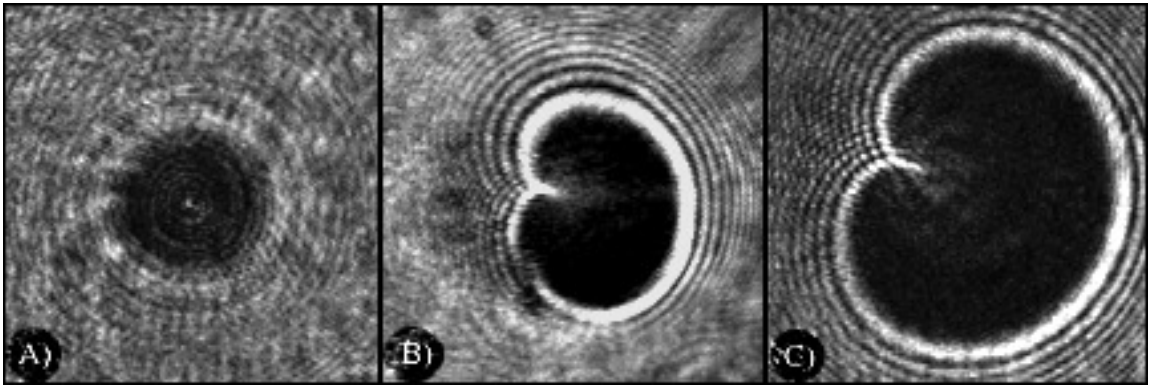

Figure 13

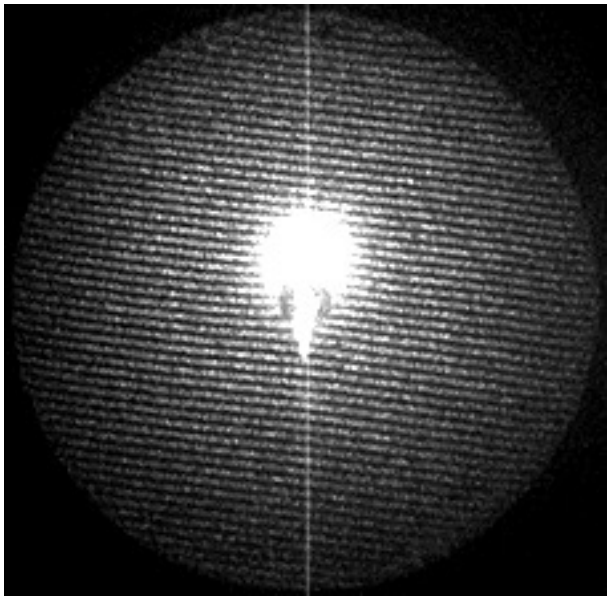

Figure 1

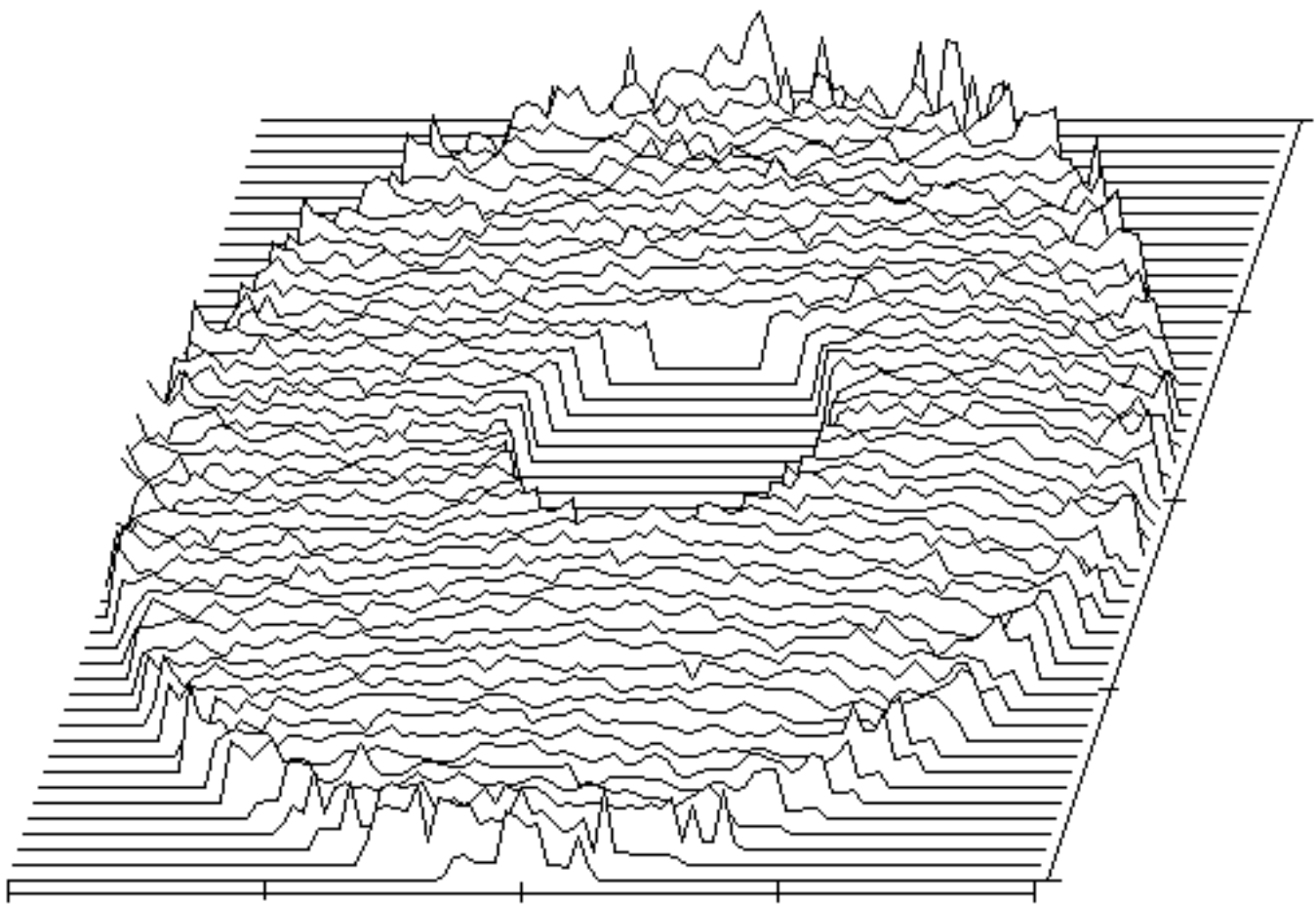

Figure 2

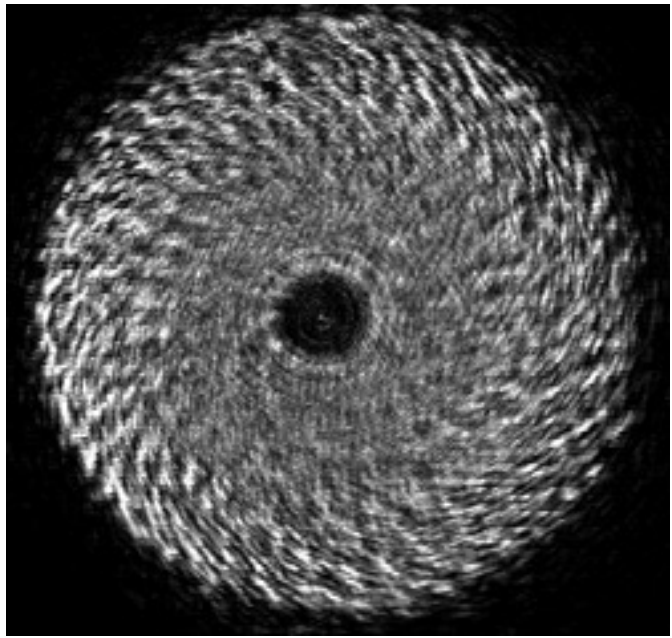

Figure 4

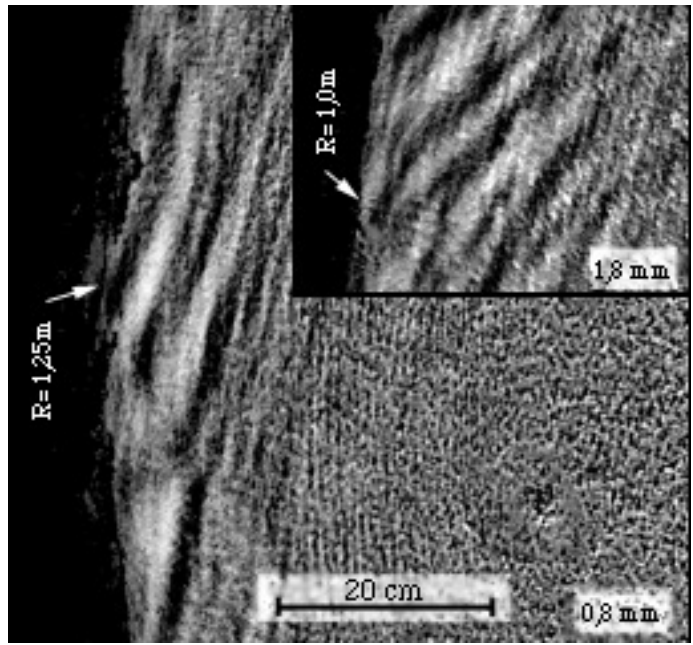

Figure 5

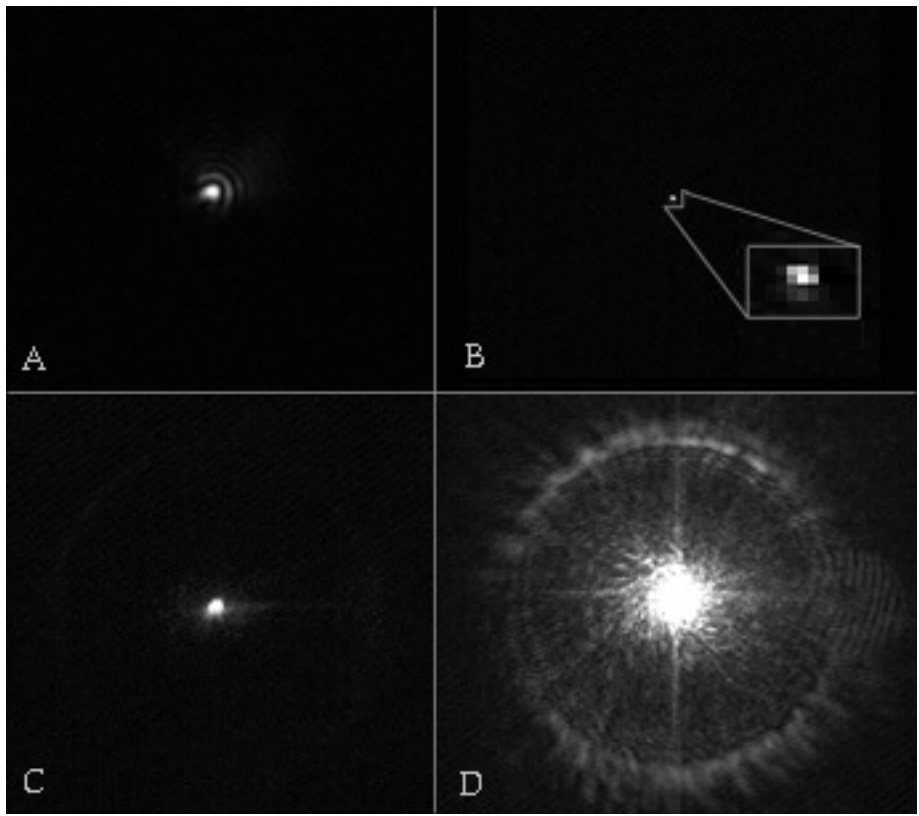

Figure 7

Figure 6

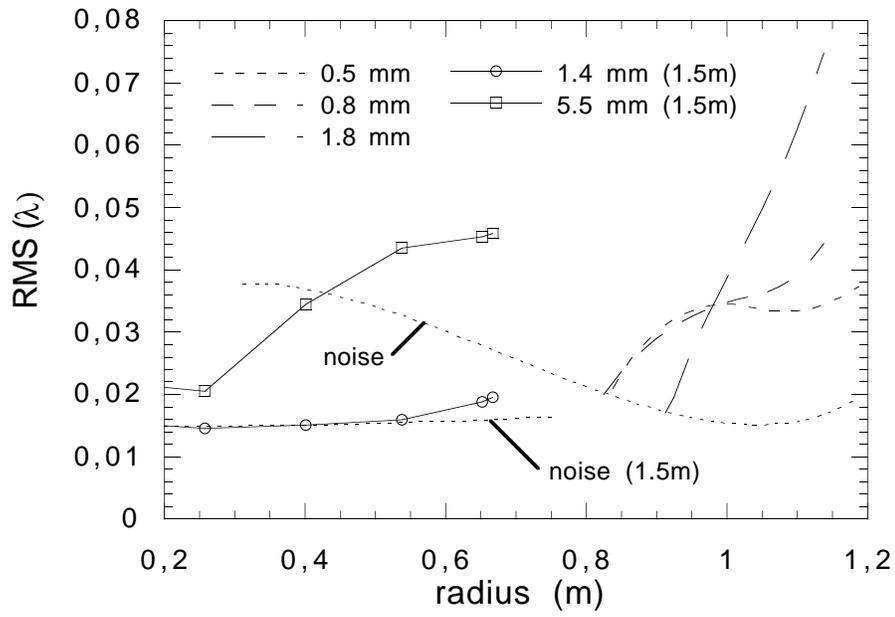

Figure 8

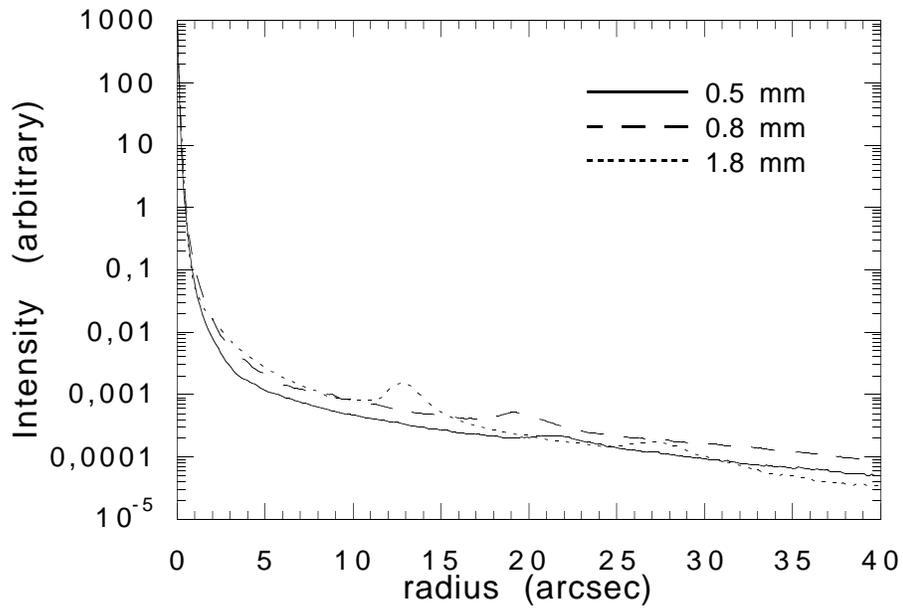

Figure 9

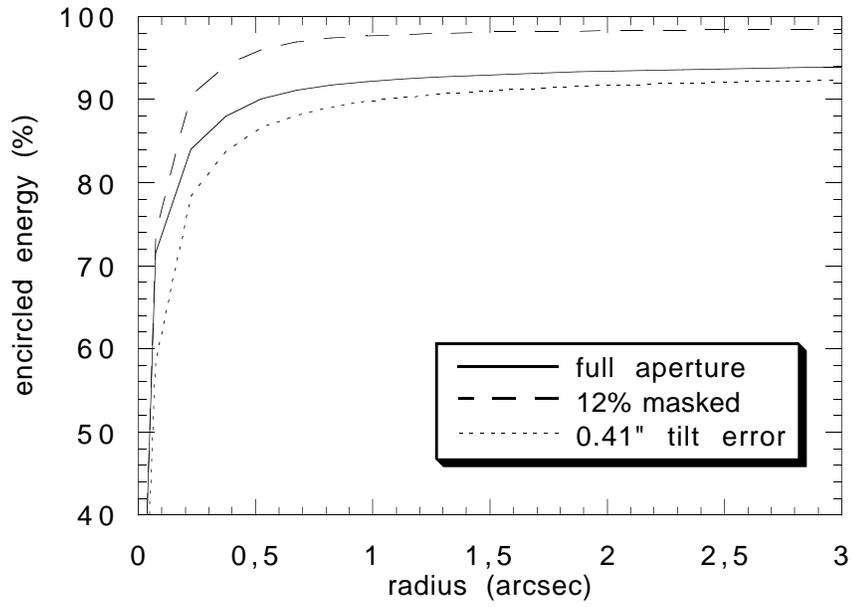

Figure 10

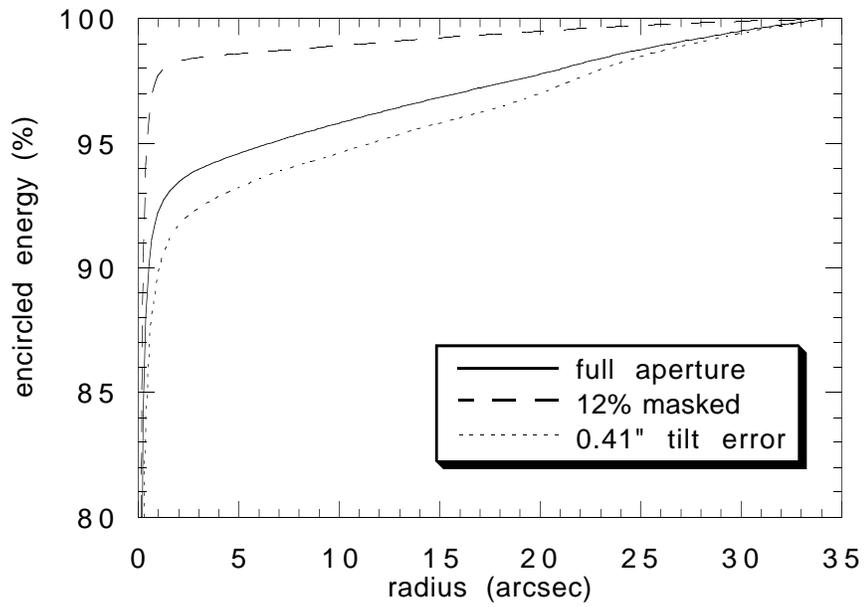

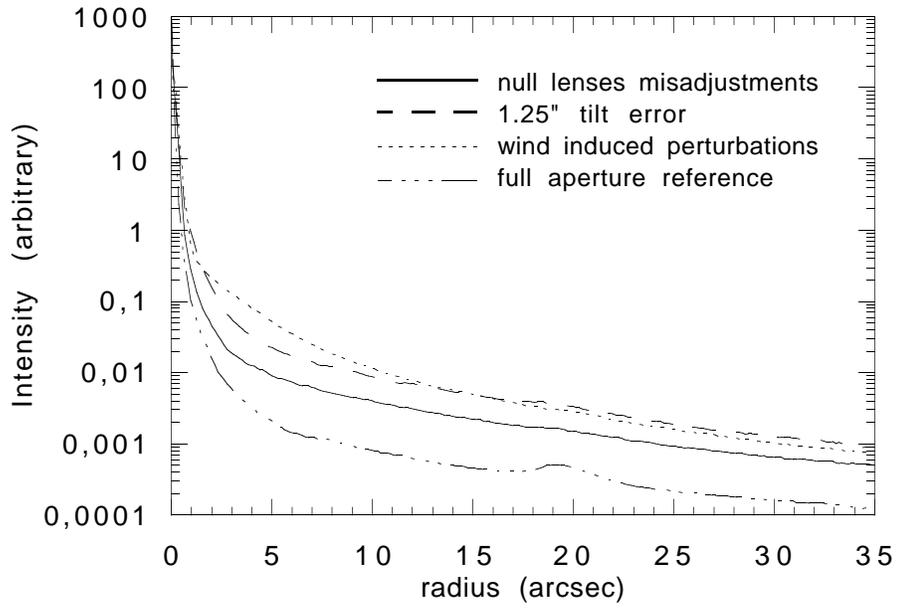
Figure 11

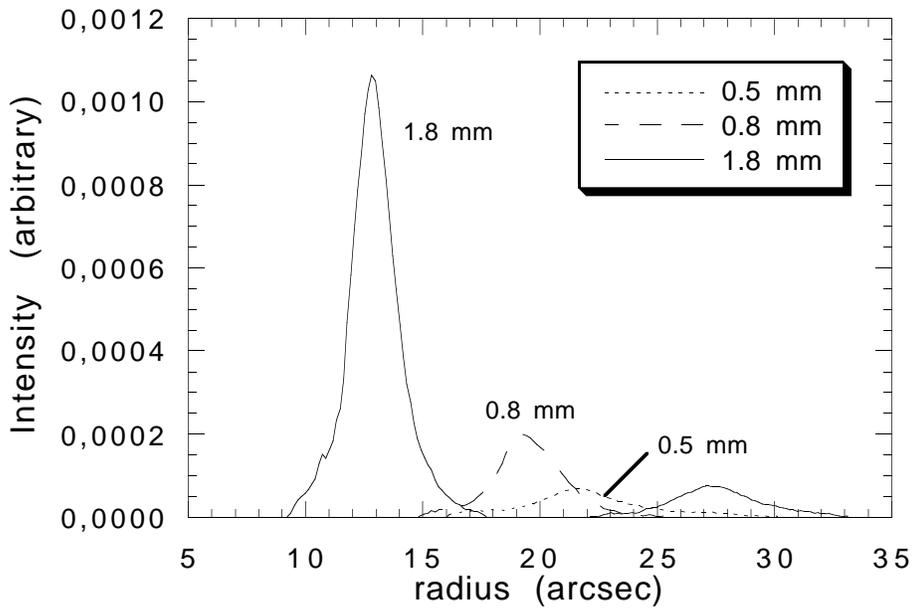
Figure 12